\documentclass[a4paper]{article}

\usepackage{hyperref}
\usepackage{graphicx}
\usepackage[numbers,square,sort&compress]{natbib}

\hypersetup{%
  backref,
  colorlinks=true,
  pdftitle={Light dark forces at flavor factories},
  pdfauthor={Luca Barze}}

\title{Light dark forces at flavor factories}

\author{L Barz\`e$^{1, 2}$\\
\footnotesize{$^1$ Dipartimento di Fisica Nucleare e Teorica,}\\\footnotesize{Universit\`a di Pavia, via A Bassi, 6, 27100, Pavia, Italy}\\
\footnotesize{$^2$ INFN, Sezione di Pavia,}\\\footnotesize{via A Bassi, 6, 27100, Pavia, Italy}\\
\small{\texttt{\href{mailto:luca.barze@pv.infn.it}{luca.barze@pv.infn.it}}}}

\date{}

\begin{document}
\maketitle

\begin{abstract}
	SuperB experiment could represent an ideal environment to test a new $U(1)$ symmetry related to light dark forces candidates. A promising discovery channel is represented by the resonant production of a boson $U$, followed by its decay into lepton pairs. Beyond approximations adopted in the literature, an exact tree level calculation of the radiative processes $e^+ e^- \rightarrow \gamma, U \rightarrow \mu^+ \mu^- \gamma, e^+ e^- \gamma$ and corresponding QED backgrounds is performed, including also the most important higher-order corrections. The calculation is implemented in a release of the generator BabaYaga@NLO useful for data analysis and interpretation. The distinct features of $U$ boson production are shown and the statistical significance is analysed.
\end{abstract}

\section{Introduction and theoretical framework}
\label{sec:intro}
The aim of this paper, which is based on \citep{Barze2010}, is to illustrate Dark Matter models which imply the existence of a new vector boson, which carries 
a new \emph{dark} force and which is lightly coupled to the photon. In particular a Monte Carlo event generator useful to describe 
this kind of light dark forces signatures at leptonic colliders  will be presented and potentiality of a SuperB experiment for this kind of studies will be shown.

The existence of an abelian gauge simmetry with an associated light $U$ boson which can interact with a really 
small coupling with Standard Model ($SM$) particles has been proposed by a wide class of new physics models\citep{Bhm2004,Pospelov2008,Pospelov20092,Cholis2009,Arkani-Hamed2009}. 

An important phenomenological support to this class of models comes in recent years, when standard astrophysics 
and particle physics seem fail to explain striking experimental signals in terms of known sources. 

These signals are represented by the 511 KeV gamma-ray signal from the galactic center observed by the INTEGRAL 
satellite\citep{Jean2003}, the excess in the cosmic ray positrons reported by PAMELA\citep{Adriani2009}, the total electron and positron flux 
measured by ATIC\citep{Chang2008}, Fermi\citep{Abdo2009}, and HESS\citep{Aharonian2008}, the annual modulation of the DAMA/LIBRA signal\citep{Bernabei2008} and the features 
of the low-energy spectrum of rare events reported by the CoGeNT collaboration\citep{Aalseth2008,Mambrini2010}.

If a new secluded gauge sector under which the $SM$ particles remain uncharged is included into the theoretical 
description, these evidences can be comprehensively interpreted in terms of WIMP Dark Matter ($DM$) particles 
connected to $SM$ ones with interaction terms varying from model to model. 

The simplest assumption\citep{Arkani-Hamed2009} is to add an extra $U(1)$ symmetry to $SM$ symmetry group which describe a new \emph{dark} 
force and suppose this force as carried by a new vector boson ($U$). It is possible to suppose also that $U$ boson 
can communicate with the $SM$ through a kinetic mixing term of the form

\begin{equation}
	\mathcal{L}_{mix}=\frac{\varepsilon}{2}F_{\mu\nu}^{\mathrm{em}}F_{\mathrm{dark}}^{\mu\nu}\mathrm{,}
\end{equation}

describing the interaction of the $U$ boson with $SM$ photon. This mixing term could occour through loop effects 
due to really massive WIMPs both coupled to ordinary photon and $U$ boson. In this case the $\varepsilon$ parameter 
should be lower than about $10^{-2}-10^{-3}$.

With reactions involving WIMPs in the initial state and standard particles as positrons in the final state and supposing these reactions mediated
by the $U$ boson it would be possible to explain the experimental signals described above. Since no astrophysical data 
involves anomalous production of antiproton, it is necessary to require the $U$ boson mass ($M_U$) to be lower than the 
mass of two protons. 

An interesting consequence of the existence of such a light $U$ boson is that it can be directly produced in a controlled 
environment, such as fixed target experiments \citep{Bjorken2009,Freytsis2009,Essig2009} or high-luminosity $e^+$ $e^-$ colliders at the GeV scale (flavor 
factories)\citep{Borodatchenkova2006,Batell2009,Bossi2009}.

At flavor factories, e.g. at DA$\Phi$NE, BESIII and present and future B-factories, a particularly clean and simple channel, 
insensitive to the details of how the $U$ boson takes a mass, is surely represented by the resonant radiative 
production of a $U$ boson, followed by its decay into a lepton - antilepton pair (see Fig. 1). 

\begin{figure}

	\centering
	\includegraphics[width=.4\textwidth]{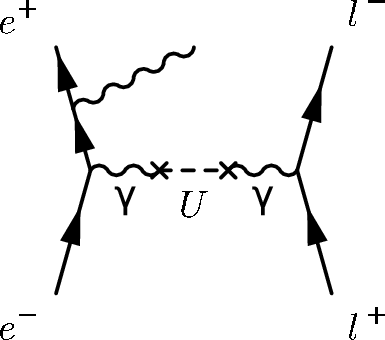}
	\caption{Typical reaction useful at leptonic colliders to study $U$ boson properties.}

\end{figure}

A distinctive feature of the expected signal is the appearance of a Breit-Wigner peak in the shape of the invariant mass 
distribution of the lepton pairs induced by the mechanism of photon radiative return and corresponding to $U$ boson resonant 
production. The drawback of this channel is the fairly small value, over a wide range of parameters, of the signal cross 
section in comparison with the large rate of the backgrounds given by the QED radiative processes 
$e^+ e^- \rightarrow l^+ l^- \gamma$, $l = e, \mu$. The signal over background ratio can be enhanced by cutting on the 
invariant mass of the lepton pair.

The conclusion presented so far in the literature is that the $U \gamma$ production process allows to identify a $U$ boson at
present flavor factories if its mass is in the range $0.1 - 1$ GeV and the kinetic mixing parameter $\varepsilon$ is greater than 
$\sim 10^{-3}$.

These analyses are generally based on the evaluation of the number of signal events through the calculation of the 
differential cross section of the $2 \rightarrow 2$ process $e^+ e^- \rightarrow U \gamma$, including the decay of the 
on-shell $U$ boson into lepton pairs by means of branching ratios\citep{Borodatchenkova2006,Reece2009}, and/or an approximate estimate of the 
backgrounds. More importantly, all the studies so far performed neglect the contribution of higher-order initial and final 
state QED corrections, which are known to be a phenomenologically relevant effect at GeV-scale $e^+ e^-$ colliders\citep{Actis2010}.

To investigate whether and how the above approximations affect the physical observables of experimental interest, a Monte Carlo
event generator, which is still missing for such studies, has been made available\footnote{you can find the code on our website (\href{http://www.pv.infn.it/~hepcomplex}{\texttt{http://www.pv.infn.it/$\sim$hepcomplex}}).} by improving the existng BabaYaga event generator for QED processes at flavor factories.

The generator can be useful for data analysis at flavor factories to study every physical observable, i.e. invariant mass 
distributions but also angular distributions or correlations and so on.

It includes:
\begin{itemize}
	\item an exact tree-level calculation of the signal and background processes contributing to the signatures 
$e^+ e^- \rightarrow \gamma, U \rightarrow \mu^+ \mu^- \gamma, e^+ e^- \gamma$;
	\item the effects of the most important higher-order corrections induced by multiple photon radiation and vacuum polarization.
\end{itemize}
We used our calculation in order to assess its impact on the experimental sensitivity as evaluated in the
literature, and show how this can be enhanced by means of event selections not considered so far. 

The calculation has been done by means of the ALPHA\citep{Caravaglios1995} algorithm, a tool to compute tree-level matrix elements. We 
implemented the calculation in the BabaYaga@NLO event generator\citep{Actis2010,CarloniCalame2000,BALOSSINI2006}, 
to simulate distributions of experimental interest and account for realistic event selection criteria. In the $U$ boson propagator 
we include the total dark photon width using the formulae of Ref. \citep{Batell2009} for the partial widths into leptons and hadrons, i.e.

\begin{equation}
	\Gamma_{U\rightarrow f^+f^-} = \frac{1}{3}\alpha\varepsilon^2 M_U \beta_f \left(1+\frac{2m_f^2}{M_U^2}\right) R(S=M_U^2)
\end{equation}

where $\beta_f = \sqrt{1-\frac{4m_f^2}{M_U^2}}$, $R=1$ for leptonic decays of the $U$ boson and
$R=\sigma_{e^+e^-\rightarrow \mathrm{hadrons}}/\sigma_{e^+e^-\rightarrow \mu^+\mu^-}$ for hadronic decays\citep{Teubner2010}.
It is worth to note that the $U$ boson is a dramatically tiny resonance, being is width well under any experimental sensitivity, varying from $10^{-7}$ to 
$10^{-2}$ for reasonable values of $M_U$ and $\varepsilon$.

It is well known that at flavor factories multiple soft and collinear radiation emitted by the colliding beams may have a strong impact on 
the measured cross section and on the shape of the distributions. The effect of higher-order corrections in our Monte Carlo generator
is taken into account using the popular QED structure function approach\citep{Nicrosini1987}. Initial state radiation modifies the tree-level cross 
section as follows

\begin{equation}
	\mathrm{d}\sigma(s,t)=\int_0^1\mathrm{d}x_1 \mathrm{d}x_2 \mathrm{d}y_1 \mathrm{d}y_2 \mathrm{d}\sigma_0 D^e(x_1,s) D^e(x_2,s) D^l(y_1,\hat{s}) D^l(y_2,\hat{s})
	\label{eq:sigmacor}
\end{equation}

where $\hat{s}, \hat{t}$ are the Mandelstam invariants after photon radiation and $D(x, s)$ is the electron structure function, which describes 
the probability to find an electron with a momentum fraction $x$ inside an electron of a given momentum $s$. If we take into account only 
the most important terms, due to resummed multiple soft and hard photon emission in the collinear approximation ($\mathcal{O}(\alpha)$), the structure function has the form:

\begin{eqnarray}
	D(x,s) & = & \frac{e^{\frac{\beta}{2}\left(\frac{3}{4}-\gamma_E\right)}}{\Gamma\left(1+\frac{\beta}{2}\right)}\frac{\beta}{2}(1-x)^{\frac{\beta}{2}-1}-\frac{\beta}{4}(1+x)\nonumber\\
	& + &\frac{\beta^2}{32}\left[(1+x)(-4 \ln(1-x)+3\ln x) - 4\frac{\ln x}{1-x} - 5 - x \right]
\end{eqnarray}

where $\beta = 2\alpha/\pi (\ln(s/m_f^2) - 1)$, $\Gamma$ is the Euler gamma function and $\gamma_E$ is the Euler - Mascheroni constant.

In Eq. \ref{eq:sigmacor} $D^e(x_i,s)$ refers to the structure functions of the initial state electron and $D^l(y_i,\hat{s})$ refers to final state
fermions.

Being an effect of size comparable to that of photon radiation, we also consider in our calculation the running of the electromagnetic coupling 
constant according to (see Ref. \citep{Actis2010} for a recent review)

\begin{eqnarray}
	\alpha(q^2) = \frac{\alpha}{1-\Delta\alpha(q^2)}\nonumber\\
	\Delta\alpha(q^2) = \Delta\alpha_l(q^2) + \Delta\alpha_h(q^2)
\end{eqnarray}

where $\Delta\alpha_l$ is the contribution due to leptons, which is analitically known, and $\Delta\alpha_h$ is the non-perturbative hadronic
contribution, included according to Ref. \citep{Teubner2010} or \citep{Jegerlehner2006}.

Fig. 2 illustrates a typical signal, the invariant mass distribution of the muon pairs for three values of $\varepsilon$ and $M_U$ 
at DA$\Phi$NE energies ($s = 1.02$ GeV), as obtained with our calculation in the lowest order approximation. It is possible to see a peak in the
distribution due to the $U$ boson effects.

\begin{figure}

	\centering
	\includegraphics[width=.6\textwidth]{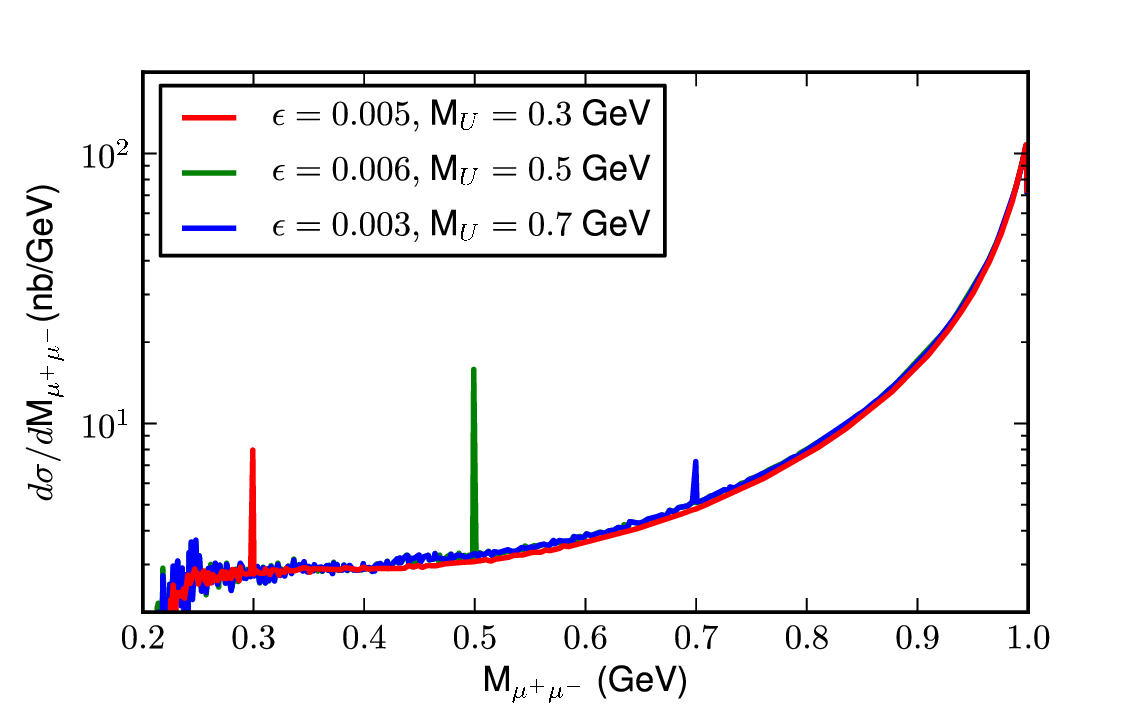}
	\caption{ Invariant mass distribution of the muon pairs for
	different values of $\varepsilon$ and $M_U$ with $s$ = 1.02 GeV}

\end{figure}

\section{Experimental sensitivity}
If we saw a peak in a distribution, for example the one related to the invariant mass of final state fermions, we would 
be able to describe the $U$ boson properties like its coupling and its mass. Instead if we couldn't see nothing but the $SM$ background, we would exclude
a certain region in the $M_U$ - $\varepsilon$ parameter space. So we need to calculate which are the $U$ boson properties that made possible to confuse 
the $U$ boson's peak with a statistical fluctuation of the background for a certain experiment with its experimental resolution and its luminosity 
(\emph{experimental sensitivity}).

On the grounds of the calculation described in Section \ref{sec:intro}, we revisited the experimental sensitivity to a dark force signal evaluated in the literature 
according to the approximations previously discussed. For concreteness, we consider the case of the KLOE/KLOE-2 experiment at the upgraded DA$\Phi$NE \citep{amelino-camelia2010} and of 
future experiments as SuperB factories\citep{Collaboration2007,Golob2010}. We also made a comparison with BaBar's analysis made by \citep{Reece2009} and the results totally agree, which means
that other calculations already present in the literature are a really good estimate of the exact calculation and that the effects due to the finite width of
the $U$ boson are negligible for the invariant mass distribution. Neverthless the Monte Carlo event generator can be useful for the experimental analysis and for 
studies of backward-forward asymmetry (which is a really smaller signal with respect to the presence of the peak in the invariant mass distribution).

We define the statistical significance ($S$) as 
\begin{equation}
	S=\frac{N_S}{\sqrt{N_B}}=\sqrt{L}\frac{(\sigma_F-\sigma_B)}{\sqrt{\sigma_B}}
\end{equation}
where $\sigma_F$ is the full cross section including the exchange of a virtual photon and $U$ boson, $\sigma_B$ ($N_B$) is the $SM$ cross section 
(number of events) background, $N_S$ is the expected number of events due to the presence of $U$ boson and $L$ is the integrated luminosity.
In the above equation we require $S$ to be greater than 5 to claim a discovery.

We assume $L = 5$ fb$^{-1}$ for KLOE/KLOE-2 and $L= 100$ ab$^{-1}$ for a B SuperB factory, respectively.
To simulate detector acceptances, we also impose the following energy and angular cuts:
\begin{description}
	\item[KLOE] $35^{\mathrm{o}} \leq \theta_{l^\pm,\gamma}\leq 145^{\mathrm{o}}$, $E_{l^\pm,\gamma}^{min} = 10$ MeV
	\item[SuperB] $30^{\mathrm{o}} \leq \theta_{l^\pm,\gamma}\leq 150^{\mathrm{o}}$, $E_{l^\pm,(\gamma)}^{min} = 30,(20)$ MeV
\end{description}
using as c.m. energies $\sqrt{s} = 1.02, 10.56$ GeV, respectively. To improve $S$, signal events can be detected as peaks in the lepton pair invariant mass 
close to the value $M_U$, in a window $M_U \pm \delta_M$. We want $\delta_M$ to be as small as possible, optimally coinciding with the detector resolution, 
a crucial parameter for these studies.

As detector resolutions we use $\delta_M = \pm 1$ MeV for KLOE/KLOE-2 and the values obtained according to the empirical relations of Ref. \citep{Reece2009} 
for B/SuperB factories, giving $\delta_M \sim \pm [1 - 10]$ MeV for a mass $M_U$ in the range 0.1 - 5 GeV.

In Fig. 3 we show the reach potential of KLOE/KLOE-2 experiment, where the sensitivity to the kinetic mixing parameter $\varepsilon$ is shown as a 
function of $M_U$. Actually, we observed that corrections affect with about the same amount both the signal and the background cross sections; 
hence the systematics induced by photon radiation largely cancel in the experimental sensitivity and the conclusions in the presence of QED radiative 
corrections confirm the ones obtained in the lowest order approximation.

\begin{figure}[h]
	\centering
	\begin{minipage}{0.47\linewidth}
	\includegraphics[width=\textwidth]{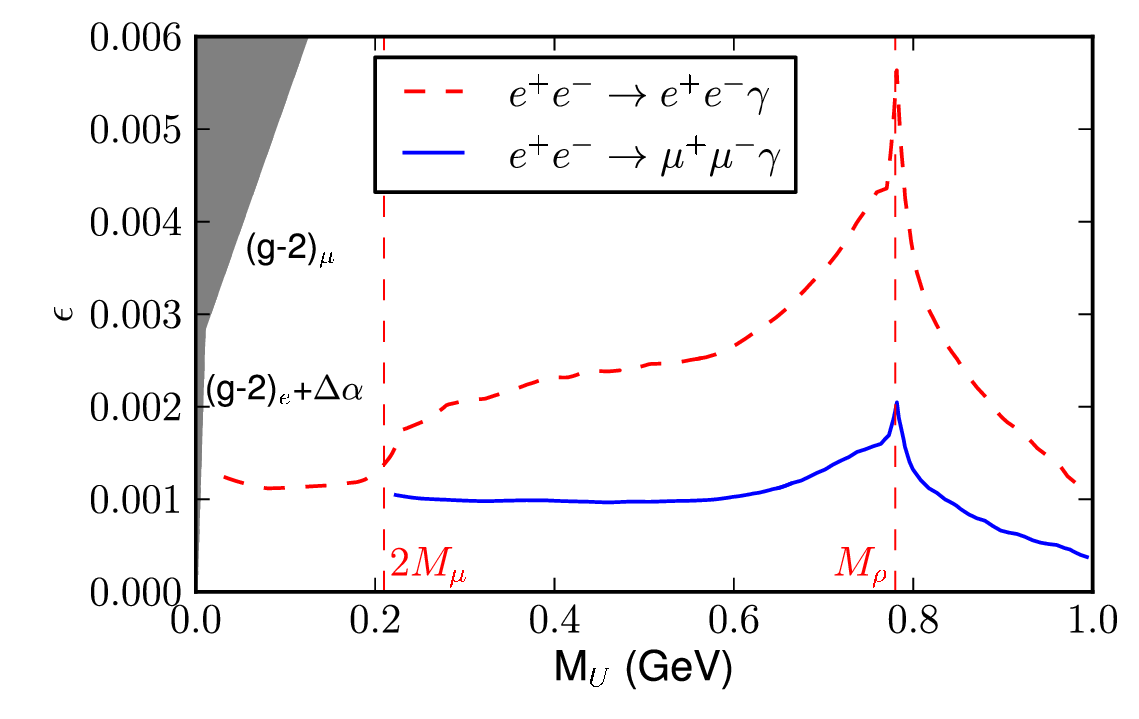}
	\caption{Experimental sensivity for KLOE/KLOE-2}
	\end{minipage}
	\hspace{0.5cm}
	\begin{minipage}{0.47\linewidth}
	\includegraphics[width=\textwidth]{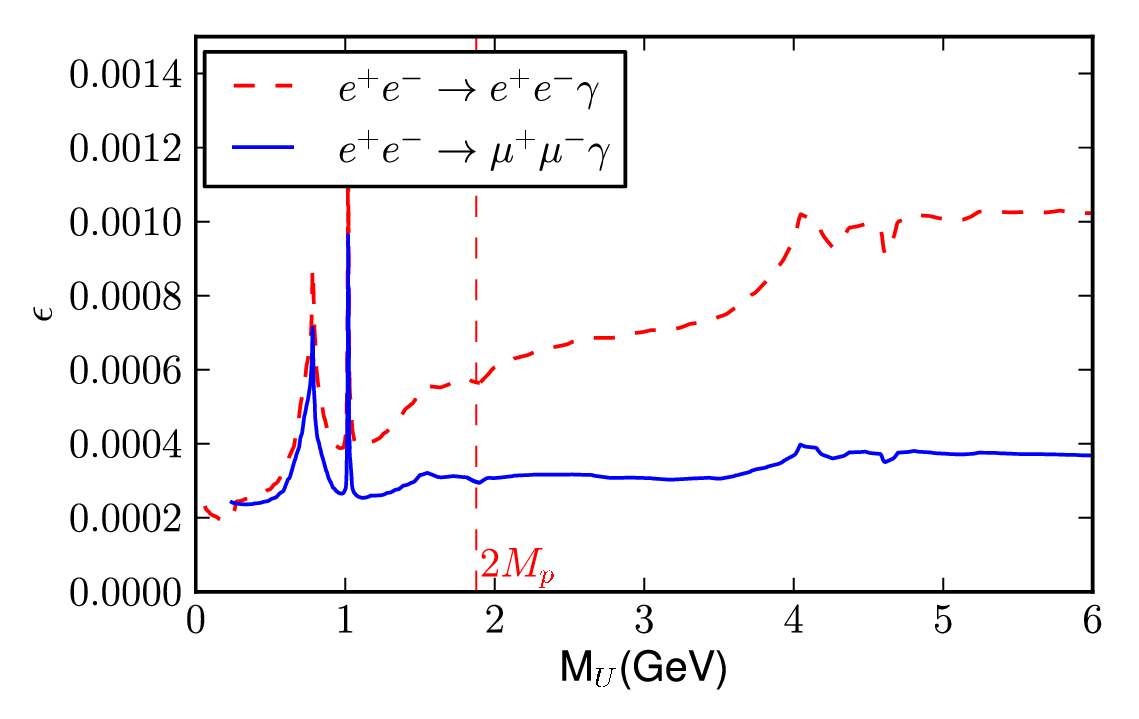}
	\caption{Experimental sensivity for SuperB}
	\end{minipage}
\end{figure}

Exclusion limits imposed by the precision measurements of the anomalous magnetic moment of the leptons and of $\alpha_{QED}$\citep{Pospelov20091} are indicated
as gray areas. 

The most striking features which could be seen in Fig. 3 are the following:
\begin{itemize}
	\item the muon channel has a better reach than the $e^+ e^-$ channel, obviously because of the smaller background. The $e^+ e^-$ channel
could instead be interesting for low values of $M_U$, but an experimental difficulty, due to the fact that it is really hard to distinguish a low energy electron
with a photon, exists and limits this kind of analysis. If we want to study a $U$ boson lighter than the $\mu^+\mu^-$ threshold we have to use some different techniques;
	\item the sensitivity is significantly degraded if $M_U$ is around the $\varrho$ resonance because the branching fraction $U \rightarrow l^+ l^-$ is 
suppressed by the dominant decay mode $U \rightarrow \pi^+ \pi^-$, which would be a more convenient channel in this region.
\end{itemize}

We have noticed that the maximum sensitivity achievable at the upgraded DA$\Phi$NE with a luminosity $L \simeq 5$ fb$^{-1}$ , i.e.
$\varepsilon \sim 0.001 - 0.002$, is equivalent to the one of the present B-factory experiments BaBar/Belle with $L \simeq 500$ fb$^{-1}$\citep{Reece2009}. This fact could be
easily understood since the reach on $\varepsilon$ follows the rule $\varepsilon^2 \propto E/\sqrt{L}$.

In our analysis we realized that radiative corrections do not significatively alter the experimental sensitivity because both the background and the signal are modified
by the same amount.

Fig. 4 presents the experimental sensitivity at a SuperB factory, where it could be possible to exclude 
an epsilon greater than few per 10$^{-4}$ for all mass values up to 2 proton mass. Hence the really high statistics of a SuperB collider will allow to probe
values of $\epsilon$ about an order of magnitude smaller than those reachable by present flavor factories.

Finally, Figs. 5 and 6 show the sensitivity for the small angle selection, defined in Eq. \ref{eq:SA}, for the KLOE/KLOE2 experiment.

\begin{eqnarray}
	\label{eq:SA}
	35^{\mathrm{o}}\leq \theta_{l^\pm} \leq 145^{\mathrm{o}}, E_{l^\pm}^{min} = 10 \mathrm{MeV}\nonumber \\
	|\cos(\theta_\gamma)|\geq \cos(15°)
\end{eqnarray}

The small angle selection is also compared with the large angle selection previously discussed. It is possible to see that there is a relevant gain in sensitivity
for $M_U$ greater than $\sim 0.5$ GeV.

\begin{figure}
	\centering
	\begin{minipage}{0.47\linewidth}
	\includegraphics[width=\textwidth]{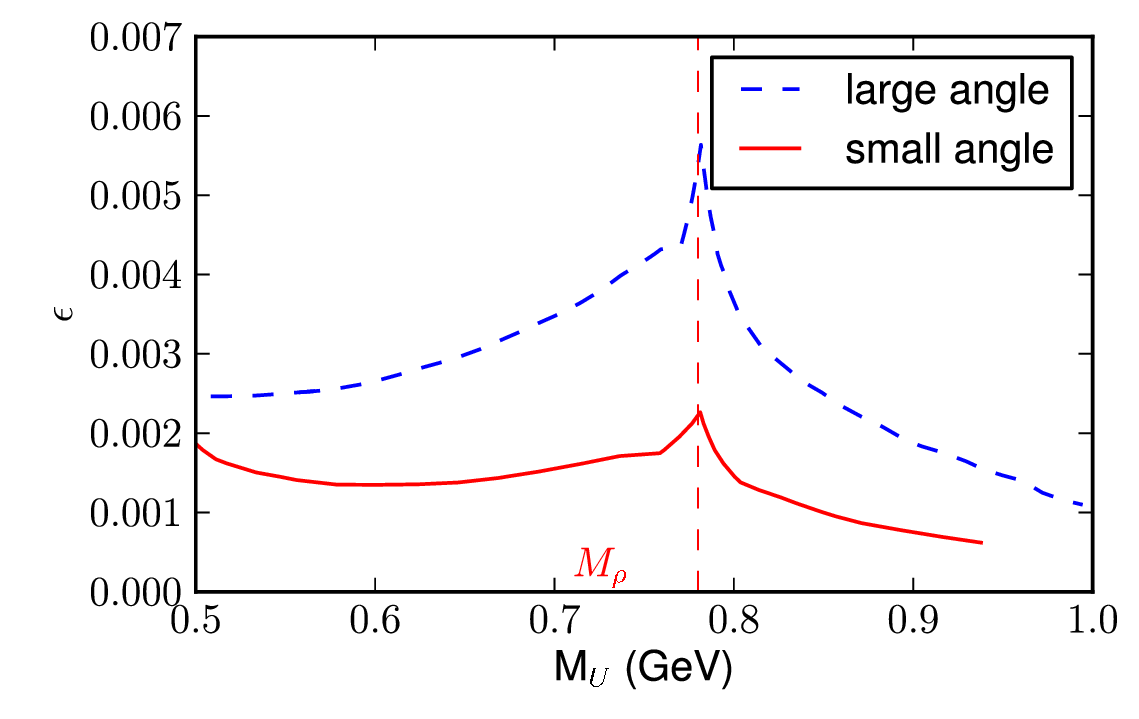}
	\caption{Experimental sensivity for KLOE/KLOE-2 - \emph{Small Angle} selection - $\mu$ channel}
	\end{minipage}
	\hspace{0.5cm}
	\begin{minipage}{0.47\linewidth}
	\includegraphics[width=\textwidth]{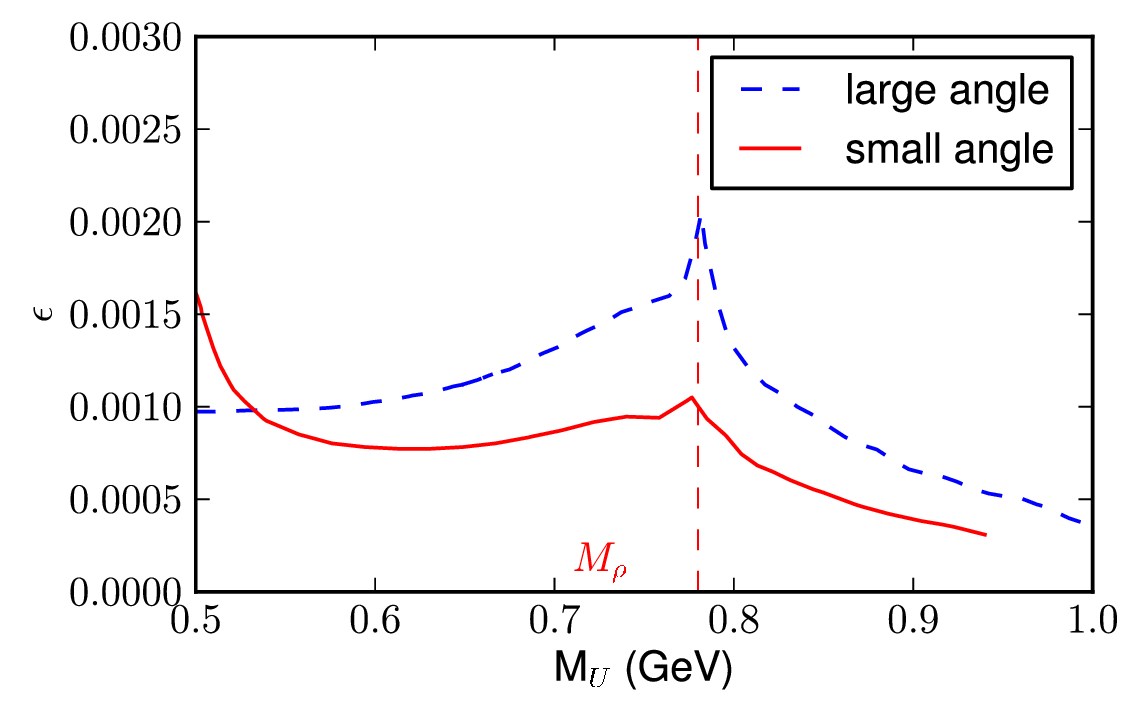}
	\caption{Experimental sensivity for KLOE/KLOE-2 - \emph{Small Angle} selection - $e$ channel}
	\end{minipage}
\end{figure}

\section{Conclusion}
With existing experiments and analysis of existing data we can exclude a $U$ boson with an $\epsilon$ coupling of about 10$^{-3}$ for a wide range of mass 
(flavor factories\citep{Barze2010}) and some particular area in the $\epsilon$ - $M_U$ parameter space for a really light $U$ boson (beam dump 
experiment\citep{Pospelov20092}). With a SuperB factory we would be able to see a $U$ boson if its kinetic mixing is above few 10$^{-4}$. Anyway a really
big area in the $\varepsilon - M_U$ space would be allowed also after the analysis of a SuperB experiment. A $U$ boson with such properties would be able
to explain the astrophysical data. To cover this area it would be useful a new fixed target experiment\citep{Essig2011} in which a really high luminosity 
could be reached.

\bibliographystyle{plainnat}
\bibliography{library}

\providecommand{\newblock}{}
\begin{thebibliography}{10}
\expandafter\ifx\csname url\endcsname\relax
  \def\url#1{{\tt #1}}\fi
\expandafter\ifx\csname urlprefix\endcsname\relax\def\urlprefix{URL }\fi
\providecommand{\eprint}[2][]{\url{#2}}

\bibitem{Barze2010}
Barz\`{e} L, Balossini G, Bignamini C, Calame C~M~C, Montagna G, Nicrosini O
  and Piccinini F 2010   6 \urlprefix\url{http://arxiv.org/abs/1007.4984}

\bibitem{Bhm2004}
Boehm C and Fayet P 2004 {\em Nuclear Physics B\/} {\bf 683} 219--263 ISSN
  05503213 \urlprefix\url{http://arxiv.org/abs/hep-ph/0305261}

\bibitem{Pospelov2008}
Pospelov M, Ritz A and Voloshin M 2008 {\em Physics Letters B\/} {\bf 662}
  53--61 ISSN 03702693 \urlprefix\url{http://arxiv.org/abs/0711.4866}

\bibitem{Pospelov20092}
Pospelov M 2009 {\em Physical Review D\/} {\bf 80} 14 ISSN 1550-7998
  \urlprefix\url{http://arxiv.org/abs/0811.1030}

\bibitem{Cholis2009}
Cholis I, Dobler G, Finkbeiner D~P, Goodenough L and Weiner N 2009 {\em
  Physical Review D\/} {\bf 80} 28 ISSN 1550-7998
  \urlprefix\url{http://arxiv.org/abs/0811.3641}

\bibitem{Arkani-Hamed2009}
Arkani-Hamed N, Finkbeiner D~P, Slatyer T~R and Weiner N 2009 {\em Physical
  Review D\/} {\bf 79} 22 ISSN 1550-7998
  \urlprefix\url{http://arxiv.org/abs/0810.0713}

\bibitem{Jean2003}
Jean P, Knodlseder J, Lonjou V, Allain M, Roques J~P, Skinner G~K, Teegarden
  B~J, Vedrenne G {\em et~al.\/} 2003 {\em Astronomy and Astrophysics\/} {\bf
  407} L55--L58 ISSN 0004-6361
  \urlprefix\url{http://arxiv.org/abs/astro-ph/0309484}

\bibitem{Adriani2009}
Adriani O, Barbarino G~C, Bazilevskaya G~A, Bellotti R, Boezio M, Bogomolov
  E~A, Bonechi L, Bongi M {\em et~al.\/} 2009 {\em Nature\/} {\bf 458} 607--9
  ISSN 1476-4687 \urlprefix\url{http://arxiv.org/abs/0810.4995}

\bibitem{Chang2008}
Chang J, Adams J~H, Ahn H~S, Bashindzhagyan G~L, Christl M, Ganel O, Guzik T~G,
  Isbert J {\em et~al.\/} 2008 {\em Nature\/} {\bf 456} 362--5 ISSN 1476-4687
  \urlprefix\url{http://dx.doi.org/10.1038/nature07477}

\bibitem{Abdo2009}
Abdo A, Ackermann M, Ajello M, Atwood W, Axelsson M, Baldini L, Ballet J,
  Barbiellini G {\em et~al.\/} 2009 {\em Physical Review Letters\/} {\bf 102} 6
  ISSN 0031-9007 \urlprefix\url{http://arxiv.org/abs/0905.0025}

\bibitem{Aharonian2008}
Aharonian F, Akhperjanian A, {Barres de Almeida} U, Bazer-Bachi A, Becherini Y,
  Behera B, Benbow W, Bernl\"{o}hr K {\em et~al.\/} 2008 {\em Physical Review
  Letters\/} {\bf 101} 5 ISSN 0031-9007
  \urlprefix\url{http://arxiv.org/abs/0811.3894}

\bibitem{Bernabei2008}
Bernabei R, Belli P, Cappella F, Cerulli R, Dai C~J, D’Angelo A, He H~L,
  Incicchitti A {\em et~al.\/} 2008 {\em The European Physical Journal C\/}
  {\bf 56} 333--355 ISSN 1434-6044
  \urlprefix\url{http://arxiv.org/abs/0804.2741}

\bibitem{Aalseth2008}
Aalseth C, Barbeau P, Cerde\~{n}o D, Colaresi J, Collar J, de~Lurgio P, Drake
  G, Fast J {\em et~al.\/} 2008 {\em Physical Review Letters\/} {\bf 101} ISSN
  0031-9007 \urlprefix\url{http://arxiv.org/abs/0807.0879}

\bibitem{Mambrini2010}
	Mambrini Y {\em JCAP\/} {\bf 2010} 09
    \urlprefix\url{http://dx.doi.org/10.1088/1475-7516/2010/09/022}

\bibitem{Bjorken2009}
Bjorken J, Essig R, Schuster P and Toro N 2009 {\em Physical Review D\/} {\bf
  80} 14 ISSN 1550-7998 \urlprefix\url{http://arxiv.org/abs/0906.0580}

\bibitem{Freytsis2009}
Freytsis M, Ovanesyan G and Thaler J 2009   38
  \urlprefix\url{http://arxiv.org/abs/0909.2862}

\bibitem{Essig2011}
Essig R, Schuster P, Toro N and Wojtsekhowski B 2010 {\em Journal of High Energy 
Physics} {\bf 2011} 2 \urlprefix\url{http://arxiv.org/abs/1001.2557}

\bibitem{Essig2009}
Essig R, Schuster P and Toro N 2010 {\em Physical Review D} {\bf 80} 1
	\urlprefix\url{http://arxiv.org/abs/0903.3941}

\bibitem{Borodatchenkova2006}
Borodatchenkova N, Choudhury D and Drees M 2006 {\em Physical Review Letters\/}
  {\bf 96} 4 ISSN 0031-9007 \urlprefix\url{http://arxiv.org/abs/hep-ph/0510147}

\bibitem{Batell2009}
Batell B, Pospelov M and Ritz A 2009 {\em Physical Review D\/} {\bf 79} 14 ISSN
  1550-7998 \urlprefix\url{http://arxiv.org/abs/0903.0363}

\bibitem{Bossi2009}
Bossi F 2009  \urlprefix\url{http://arxiv.org/abs/0904.3815}

\bibitem{Reece2009}
Reece M and Wang L 2009 {\em Journal of High Energy Physics\/} {\bf 2009}
  051 ISSN 1029-8479 \urlprefix\url{http://arxiv.org/abs/0904.1743}

\bibitem{Actis2010}
Actis S, Arbuzov A, Balossini G, Beltrame P, Bignamini C, Bonciani R, {Carloni
  Calame} C~M, Cherepanov V {\em et~al.\/} 2010 {\em The European Physical
  Journal C\/} {\bf 66} 585--686 ISSN 1434-6044
  \urlprefix\url{http://arxiv.org/abs/0912.0749}

\bibitem{Caravaglios1995}
Caravaglios F and Moretti M 1995 {\em Physics Letters B\/} {\bf 358} 332--338
  ISSN 03702693 \urlprefix\url{http://arxiv.org/abs/hep-ph/9507237}

\bibitem{CarloniCalame2000}
{Carloni Calame} C~M 2000 {\em Nuclear Physics B\/} {\bf 584} 459--479 ISSN
  05503213 \urlprefix\url{http://arxiv.org/abs/hep-ph/0003268}

\bibitem{BALOSSINI2006}
Balossini G, {Carloni Calame} C~M, Montagna G, Nicrosini O and Piccinini F 2006
  {\em Nuclear Physics B\/} {\bf 758} 227--253 ISSN 05503213
  \urlprefix\url{http://arxiv.org/abs/hep-ph/0607181}

\bibitem{Teubner2010}
Teubner T, Hagiwara K, Liao R, Martin A~D and Nomura D 2010   6
  \urlprefix\url{http://arxiv.org/abs/1001.5401}

\bibitem{Nicrosini1987}
Nicrosini O and Trentadue L 1987 {\em Physics Letters B\/} {\bf 196} 551--556
  ISSN 03702693 \urlprefix\url{http://dx.doi.org/10.1016/0370-2693(87)90819-7}

\bibitem{Jegerlehner2006}
Jegerlehner F 2006 {\em Nuclear Physics B - Proceedings Supplements\/} {\bf
  162} 22--32 ISSN 09205632 \urlprefix\url{http://arxiv.org/abs/hep-ph/0608329}

\bibitem{amelino-camelia2010}
Amelino-Camelia G, Archilli F, Babusci D, Badoni D, Bencivenni G, Bernabeu J,
  Bertlmann R~A, Boito D~R {\em et~al.\/} 2010   60
  \urlprefix\url{http://arxiv.org/abs/1003.3868}

\bibitem{Collaboration2007}
Collaboration S 2007   524 \urlprefix\url{http://arxiv.org/abs/0709.0451}

\bibitem{Golob2010}
Golob B 2010  \urlprefix\url{http://arxiv.org/abs/1006.4208}

\bibitem{Pospelov20091}
Pospelov M and Ritz A 2009 {\em Physics Letters B\/} {\bf 671} 391--397 ISSN
  03702693 \urlprefix\url{http://arxiv.org/abs/0810.1502}

\end{thebibliography}

\end{document}